\documentclass[11pt]{article}
\usepackage{erice,epsfig}

\usepackage{amsmath,amssymb}
\usepackage{latexsym}

\def\noCP{$\rlap{CP}\,\diagup$\ \ }

    \def\NIM{{\em Nucl. Instrum. Methods\ }}
    \def\NP{{\em Nucl. Phys.\ }}

    \def\PR{{\em Phys. Rev.\ }}
    
    \def\EJ{{\em Eur. Phys. Journ.\ }}
    \def\SYaf{{\em Sov. Yad. Fiz.\ }}

\begin{document}
\vspace*{4cm}
\title{PHOTON COLLIDERS IN PHYSICAL PROGRAM FOR LARGE
COLLIDERS\\{\it Talk given at Workshop on Electromagnetic Probes
of Fundamental Physics, Eriche, Sicily, 16-21 October 2001 }}

\author{I.F. Ginzburg$^{1}$, S.I. Polityko$^2$  }

\address{{\it $^1$  Institute of Mathematics and
  Novosibirsk State University,} {\it Novosibirsk, Russia;}\\
 $^2$ {\it Irkutsk State University, Irkutsk, Russia}}

\maketitle\abstracts{In this report our main attention is focused
on the problems which can be solved at Photon Colliders naturally
but are hardly solved at LHC and $e^+e^-$  Linear Collider: {\it
The New Physics --- new particles, new interactions,
supersymmetry, Dirac monopole, extra dimensions... SM and the case
of a SM--like scenario --- EWSB, Higgs bosons, anomalies in
interactions of gauge bosons, QCD, Pomeron and odderon... {\em
By-product:} Production of light Goldstone particles from region
of $e\to\gamma$ conversion.} }

\section{Introduction}

{\bf Photon Collider} will be a specific mode of Linear Collider
(LC) obtained with the aid of laser light backscattering on the
accelerated electrons in Linear Collider near the collision point
\cite{GKST,TESLAPh}. Let us enumerate its main characteristics in
the frame of TESLA project \cite{TESLA}.\vspace{-3mm}
\begin{enumerate}
\item  Characteristic photon energy $E_\gamma\approx 0.8E$
($E$ -- energy of  electron in the basic $e^+e^-$
collider).\vspace{-3mm}
\item {\it For high energy peak,} $E_{\gamma 1,2}>0.7E_{\gamma max}$
(separated well from low energy part of spectrum)\vspace{-3mm}
\begin{itemize}
\item  Luminosity ${\cal L}_{\gamma\gamma}\approx {\cal L}_{ee}/3$,
${\cal L}_{e\gamma}\approx {\cal L}_{ee}/4$ with ${\int{\cal
L}_{\gamma\gamma}dt},\;{\int{\cal L}_{e\gamma}dt}\approx 200\div
150$ fb$^{-1}$/year.\vspace{-2mm}
\item Mean energy spread $<\Delta E_\gamma>\approx 0.07E_\gamma$
(by factor $2\div 3$ worse than in $e^+e^-$ mode considering
beamstrahlung and ISR).\vspace{-2mm}
\item  Mean photon helicity $<\lambda_\gamma> \approx 0.95$, with
easily variable sign. One can also get the linear polarization.
(In the $e^+e^-$ mode only longitudinal polarization is relevant.)
\end{itemize}\vspace{-4mm}
\item  The $e\to\gamma$ conversion region is  $e\gamma$ collider with
$\sqrt{s_{e\gamma}}\approx 1.2$ MeV and ${\cal L}\sim 0.1$
fb$^{-1}$/sec!\vspace{-3mm}
\item The total additional cost is estimated as about 10\% from
that of $^+e^-$  LC.
\end{enumerate}

The Standard Model (SM) is verified now with high precision except
for mechanism of electroweak symmetry breaking (EWSB). The new
large colliders are constructed with the dream to find New Physics
and understand EWSB mechanism. We discuss {\bf different physical
pictures obtained after  running of large colliders of next
generation } (Tevatron, LHC and LC):\vspace{-3mm}

\begin{enumerate}
\item Clear signals of New Physics (new particles, strong
deviations from SM) will be found.\vspace{-3mm}

\item The physical picture will coincide with that expected in SM within
experimental precision --- {\em SM--like scenario}, determined for
the time of observations:\vspace{-3mm}
\begin{itemize}
\item No new particles and interactions will be discovered except
for single Higgs boson.\vspace{-2mm}
\item The couplings of Higgs boson to $W$, $Z$ and quarks will
coincide with those predicted in SM within the experimental
precision.\vspace{-3mm}
\end{itemize}
\end{enumerate}

We discuss the potential of Photon Collider in both these
scenarios separately. For more details on some mentioned points
see refs. \cite{TESLAPh,gold}.

\section{Hunting for New Physics}
\vspace{-3mm}
\subsection{Discovery of new particles}

The discovery of a new particle will be a clean signal of some
definite form of a new theory. In the discussion below we denote
the kinematically allowed discovery bound for its mass as $M_b$.

The production of pair of charged particles in $\gamma\gamma$
collisions near the threshold is described by QED with reasonable
accuracy. The corrections due to other (even strong) interactions
can be neglected in the estimates of opportunity to discover the
particle. Therefore, the real discovery limit of new particle in
$\gamma\gamma$ channel is close to $M_b\approx 0.8E$, it is lower
than that in $^+e^-$ mode.

The $e\gamma$  collisions provide us with final states which
cannot be produced with similar intensity other ways. In reactions
like $e\gamma\to BA$ with light particle $A$ and new particle $B$
the kinematical discovery bound $M_b$ can be much higher than in
other reactions, $M_b\le 1.8E$. However, the cross section of such
process depends on new coupling constants like $eAB$. Therefore --
in contrast to $\gamma\gamma$ collisions -- the absence of such
signals can be explained by absence of new particle $B$ as well as
by its "electrofobic" nature (very small $eAB$ coupling).
\vspace{3mm}

$\blacksquare$ {\bf\boldmath  The $e\gamma$ mode provides the best
opportunities for discovering a number of new particles.} The
discovery limits $M_b$ of some new states are presented in Table
1.
 \begin{table}[hbt]\vspace{-2mm}
\begin{center}
\begin{tabular}{|c|c|c|c|c|c|}\hline
&&name& reaction&$M<M_b$&observed\\\hline
 &$e^*$& excited $e$&$e\gamma\to
 e^*$&1.8E&$e\gamma$\ or $eZ$\\\cline{2-6}
& $\nu_e^*$& excited $\nu$& $e\gamma\to
 W\nu_e^*$&$1.8E-M_W$&$eW$\\\cline{2-6}
& $W'$&new $W$&$e\gamma\to \nu W'$& 1.8E&$WZ$,
$W\gamma$\\\hline\hline \vspace{-4mm}&&&&&\\
 S& $\tilde{W}$& wino& $e\gamma\to
 \tilde{W}\chi$&$1.8E-M_\chi$&\\\cline{2-6} \vspace{-4mm}&&&&&\\
U&$\tilde{Z}$& zino& $e\gamma\to
 \tilde{Z}\tilde{e}$&$1.8E-M_{\tilde{e}}$&\\\cline{2-6}
S&$\tilde{e}$& selectron&$e\gamma\to
\tilde{e}\chi$&$1.8E-M_\chi$&\\\cline{2-6}
 Y&\multicolumn{5}{|c|}{$\chi$ -- LSP, lightest superparticle}\\\hline
\end{tabular}
\caption{\it Some discovery limits for $e\gamma$ mode}
\end{center}\vspace{-3mm}
\end{table}
For the excited electron the expected cross section is high enough
to observe it even with the weak enough coupling constant.
\vspace{3mm}

$\blacksquare$ {\bf\boldmath  The $\gamma\gamma$ mode. Pair
production}. The cross section of the pair production
$\gamma\gamma\to P^+P^-$ ($P=S$ -- scalar, $P=F$ -- fermion, $P=W$
-- gauge boson) not far from the threshold is given by QED as
($\lambda_i$ -- circular, $\ell_i$ -- linear polarization of
photons, $\phi= \angle{(\vec{\ell}_1,\vec{\ell}_2)}$)
 \begin{equation}\begin{array}{c}
\sigma=\frac{\pi\alpha^2}{M_P^2}C_qf_P\left(\frac
{s}{4M_P^2}\right)
 \approx 65
\left(\frac{1\,TeV}{M_P}\right)^2 C_q
f_P\left(\frac{s}{4M_P^2}\right)\mbox{fb};\\ f_P(x) \equiv
f_P^0(x) +\lambda_1\lambda_2\; g_P^a +\ell_1\ell_2 \cos 2\phi\;
g_P^\tau\,;\quad C_q=\left\{\begin{array}{c c l} 1&\mbox{ if
}&\,Q=1,\\ 3Q^4&\mbox{ if }&\, Q\mbox{ is noninteger}
 \end{array}\right.
\end{array}
 \end{equation}
(with functions $f_P^0$, $g_P^a$ and $g_P^\tau$ written e.g. in
ref.~\cite{SDiego}). These $f_P^0$ (for the unpolarized photons)
are shown in Fig.~1 ($W^2=s$). The $\gamma\gamma$ cross sections
are evidently higher than the corresponding $e^+e^-$ collisions,
which are also shown here \footnote{ The observable $e^+e^-\to
P^+P^-$ cross section includes also $Z^*$ contribution, dependent
on other quantum numbers besides charge and spin.  At $s\gg M_Z^2$
typically $\sigma(e^+e^-\to P^+P^-)<1.3\sigma(e^+e^-\to
\gamma^*\to P^+P^-)$.}.
\begin{figure}[thb]
\begin{center}
 \epsfig{file=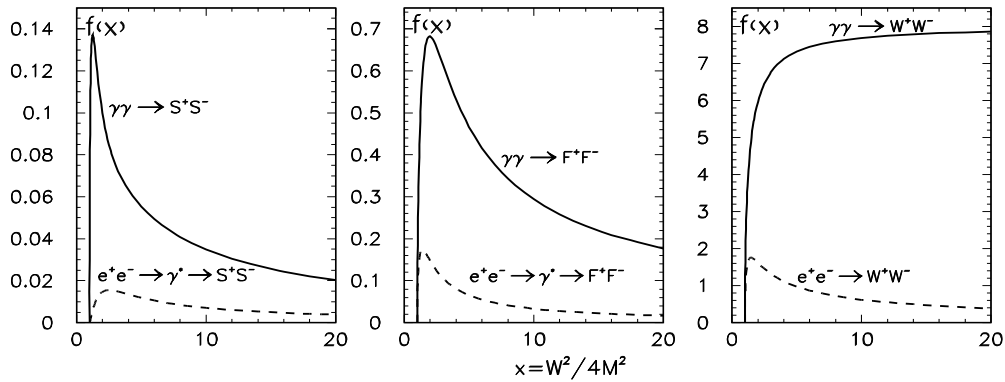,height=5cm,width=0.95\textwidth}
 \end{center}
\caption{\it The $\gamma\gamma\to P^+P^-$ cross sections, divided
by ($\pi\alpha^2/M_P^2$), nonpolarized photons, $C_q=1$. The cross
sections $e^+e^-\to \gamma^*\to P^+P^-$ are also shown. }
\end{figure}

With expected luminosities the first discovery of new particle is
preferable in $e^+e^-$ mode. After that, the key problem will be
to study the nature of the discovered particle. In this respect,
the $\gamma\gamma$ production provides essential advantages
compared to $e^+e^-$ collisions:\vspace{-3mm}
\begin{itemize}
\item These cross sections  decrease slow with energy growth.
Therefore, one can study these processes relatively far from the
threshold where the decay products don't overlap.\vspace{-3mm}
\item Near the threshold $f_P\propto (1 +\lambda_1\lambda_2
\pm\ell_1\ell_2 \cos2\phi)$\  with + sign for $P=S$ (scalar) and
-- sign for $P=F$ (fermion). This polarization dependence provides
the opportunity to know spin of produced particle independent on
its charge. (This problem arises, e.g., at the discovery of SUSY
particles since spin of invisible neutral is
unknown).\vspace{-3mm}
\item The possible CP violation in the $P\gamma$ interaction can
be seen as a variation of cross section with changing the sign of
photon helicity.
\end{itemize}

$\blacksquare$ {\bf\boldmath The  discovery of some other
particles in $\gamma\gamma$ mode}.\vspace{-3mm}
\begin{itemize}
\item {\em The leptoquark} $(\ell q)$  can be discovered in
reactions like $\gamma\gamma\to \bar{\ell}+t +(\ell t)$ with
$M_b\approx 1.5 E_\gamma$.\vspace{-3mm}
\item {\em The scalar or tensor resonances $R$ appearing due to the strong
interaction in Higgs sector} can be discovered in process
$\gamma\gamma\to R$ with $M_b\approx 2E_\gamma$.\vspace{-3mm}
\item {\em The gluino} $\tilde g$ can be produced in process
$\gamma\gamma\to \tilde{g}\tilde{g}$ (via quark loop) \cite{Li}.
The maximal value of this cross section is
$\sim(\alpha^2\alpha^2_s/ M_{\tilde{q}}^2)
\ln(M^2_{\tilde{q}}/M_q^2)$ at $2M_{\tilde{g}}<\sqrt
{s_{\gamma\gamma}}<2M_{\tilde{q}}$. For example, this cross
section is about 1 pb at $M_{\tilde{q}}=0.5$ TeV,
$M_{\tilde{g}}=0.25$ TeV for $\sqrt{s_{\gamma\gamma}}=1.5$
TeV.\vspace{-3mm}
\item If the stop squark is not too heavy, the very
narrow {\em atom-like (scalar) stoponium} with mass 200-400 GeV
should also exist (such states cannot be observed at hadron
collider). It can be clearly seen at $\gamma\gamma$  collider with
cross section averaged over photon spectrum $<\sigma>\approx
10-50$ fb and clear enough signature \cite{GorIl}.
 \end{itemize}
\vspace{-5mm}

\subsection{ Some problems with new interactions}.\vspace{-3mm}

$\bullet$ Due to high values of the basic cross sections for the
pair production of charged particles, Photon Collider would be an
excellent place for observation of (even small) {\bf possible
flavor changing neutral currents (FCNC)}, for example, with
superparticles.

$\bullet$ In the 2HDM or MSSM in the CP conserving case the masses
of heavy  Higgs scalar $H$ and pseudoscalar $A$ can be close to
each other or they are mixed (in the CP violated -- \noCP --
case). The observations of decay products at LHC and $e^+e^-$ LC
cannot resolve these opportunities due to low resolution for these
bosons. The polarization asymmetries in Higgs boson production at
Photon Collider can resolve these variants, i.e. establish,
whether {\bf CP parity in Higgs sector is violated or not}.\\

$\blacksquare$ {\bf\boldmath Using the $\gamma\gamma\to
\gamma\gamma$, etc. processes for the  search of effects from
extra dimensions or heavy point--like Dirac monopole.} In both
cases the process is considered strongly below new mass scale $M$
or particle production threshold $2M$. In both cases the cross
section can be written as
 \begin{equation}
\sigma(\gamma\gamma\to\gamma\gamma) =\frac{A}{32 \pi s}
\left(\frac{s}{4 M^2}\right)^4 \label{mon}
 \end{equation}
with specific polarization dependence and angular distribution
($S$ and $D$ waves, roughly --- isotropic). This wide angle
elastic $\gamma\gamma$ scattering has very clear signature and
small QED background. The observation of strong elastic
$\gamma\gamma$ scattering raising quickly with energy will be the
signal of one of these mechanisms.  The study of polarization and
angular dependence at photon collider and some similar processes
can discriminate what mechanism is relevant.

$\bullet$ {\bf Effects of extra dimensions} \cite{extra} are
considered in the scenario where gravity propagates in the
$(4+n)$--dimensional bulk of space-time, while gauge and matter
fields are confined to the (3+1)--dimensional world volume of a
brane configuration. The extra $n$ dimensions are compactified
with scale $R$ what produces the Kaluza--Klein excitations having
masses $\pi n/R$. The corresponding scale in our world is assumed
to be $M\sim$\  few TeV. The particles of our world interact (as
$A{\bar A}\to B{\bar B}$) via the set of Kaluza-Klein excitations
having spin 2 or 0 as e.g. $T^{\mu\nu}T^{\mu\nu}/M^4$, where
$T^{\mu\nu}$ is stress-energy tensor. The coefficients are
accumulated in the definition of $M$ (with $A\approx 1$).

The $\gamma\gamma$ initial state has numerical advantage as
compared to $e^+e^-$ one. The $\gamma\gamma$  final state has the
best signature and the lowest SM background. The interference
(with SM) effect enhances this anomaly for $\gamma\gamma\to WW$
process (simultaneously with enhancement of background).

$\bullet$ {\bf Point--like Dirac monopole} \cite{GPS}. This
monopole existence would explain mysterious quantization of the
electric charge.  There is no place for it in modern theories of
our world but there are no convincing reasons against its
existence.

At $s\ll M^2$ the electrodynamics of monopoles is expected to be
similar to the standard QED. At $g\sqrt{s}/(4\pi M)<1$ (with
$g=n/(2e)$) the effect is described by monopole loop, and one can
use coefficient $A\propto g^8$, calculated within QED. Both $A$
and details of angular and polarization dependence depend strongly
on spin of monopole $J$, e.g., $A(J=1)/A(J=0)\approx 1900$.

Effect can be seen at TESLA500 at $M<4-10$ TeV (depending on
monopole spin). Modern limitation (Tevatron) is about 10 times
lower.\vspace{-3mm}

\section{ SM and SM--like scenario}

If after experiments at LHC and $e^+e^-$ LC a SM--like scenario
(defined in the beginning of report) will be realized, the main
problems for study at Colliders become the following:\vspace{-3mm}
\begin{itemize}
\item The study of EWSB mechanism.\vspace{-3mm}
\item Discovery of signals of New Physics via (small) deviations
from SM predictions.\vspace{-3mm}
\item Description of observed phenomena in SM, especially QCD:\\
 $\Box$ Total cross sections, Pomeron and odderon, minijets.\\
 $\Box$ Photon structure function.\\
 $\Box$ Some problems with heavy quarks.
\end{itemize}\vspace{-3mm}
Photon Colliders provide new (sometimes unique) keys for solving
all these problems.

\subsection{The study of EWSB }

Assuming Higgs mechanism for EWSB, one should consider 3 main
opportunities:

$\bullet$ SM with one Higgs doublet and $M_h<400\div 700$ GeV --
standard SM case.

$\bullet$ More complex Higgs sector, the simplest variant -- Two
Doublet Higgs Model (2HDM).\\ In these cases Higgs boson will be
discovered at the Tevatron or LHC, its spin and couplings to $W,\,
Z$ and fermions will be precisely measured at $e^+e^-$ LC.

$\bullet$ SM with one Higgs doublet but with strong Higgs
self--interaction. In this case standard Higgs particle does not
exist and strong interaction in Higgs sector will manifest itself
as the strong interaction of gauge bosons $W$ and $Z$ (their
longitudinal components).

The measuring of $h\gamma\gamma$ and $hZ\gamma$ couplings is an
excellent tool for these problems. Indeed,\vspace{-3mm}
\begin{itemize}
\item[$\Box$] In the SM these couplings appear only at the loop level.
Therefore, the S/B for new signals is better than that for the
processes allowed at tree level.\vspace{-3mm}
\item[$\Box$] All fundamental charged particles contribute to these
effective couplings.\vspace{-3mm}
\item[$\Box$] The expected accuracy in the measurement of the two-photon
width is  $\sim 2$\% at the luminosity integral 30~fb$^{-1}$ and
$M_h\le 150$ GeV \cite{Jik}. This uncertainty can be reduced with
the expected 3--year luminosity integral about 500
fb$^{-1}$.\vspace{-3mm}
\end{itemize}

$\blacksquare${\bf The anomalous interactions of Higgs boson with
light}  \cite{GIIv} (CP conserving and nonconserving) can be
summarized in an effective interaction
 \begin{equation}
hv\left(\theta_\gamma\frac{F_{\mu \nu } F^{\mu\nu}}
{\Lambda_{\gamma}^2} + 2\theta_Z \frac{Z_{\mu \nu } F^{\mu
\nu}}{\Lambda_Z^2}+ i\theta_{P\gamma}\frac{F_{\mu \nu }
\tilde{F}^{\mu\nu}} {\Lambda_{P\gamma}^2} + 2i\theta_{PZ}
 \frac{Z_{\mu \nu } \tilde{F}^{\mu \nu}} {\Lambda_{PZ}^2}\right),
\;\; \left(\theta_i=e^{i\xi_i}\right).\label{anom}
 \end{equation}
Here $F^{\mu\nu}$ and $Z^{\mu\nu}$ are standard field strengths
for the electromagnetic and $Z$ field, $\tilde{F}^{\mu\nu} =
\varepsilon^{\mu\nu\alpha\beta} F_{\alpha\beta}/2$,
$v=(G_F\sqrt{2})^{-1/2}$ -- v.e.v. of Higgs field. Finally,
$\xi_i$ are the phases of couplings, in general not equal to 0 or
$\pi$.
\begin{figure}[thb]
\centering \epsfig{file=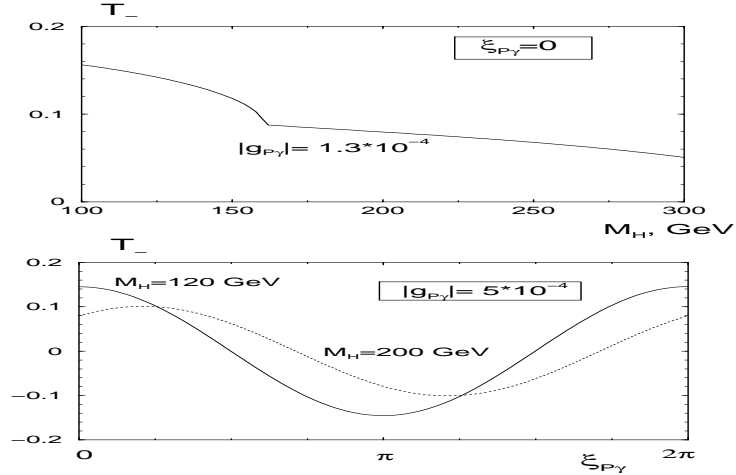,height=6cm,width=0.6\textwidth}
\caption{\it  The longitudinal asymmetry in $\gamma\gamma\to h$
production due to anomalous interactions (\ref{anom}). }
\end{figure}

The CP conserving anomalies give the deviation of measured
$\gamma\gamma\to h$ and $e\gamma\to eh$ cross sections from SM
prediction. The \noCP anomalies give the polarization asymmetries
-- variation of cross sections with change of sign of photon
helicities (longitudinal), or with the angle between directions of
linear polarization of colliding photons (transverse). One example
-- the longitudinal asymmetry -- is shown in Fig.~1. We see that
the effects can be seen at reasonable values of anomaly scales
$\Lambda_i$ and phases $\xi_i$.\\

$\blacksquare$ {\bf Distinguishing SM/2HDM in the SM--like
scenario} \cite{GKO}. The SM--like scenario can be realized both
in the SM and in other models. The simplest extension of Higgs
sector is the 2HDM with the Model II for the Yukawa coupling (the
same is realized in MSSM). It contains 2 Higgs doublet fields
$\phi_1$ and $\phi_2$ with v.e.v.'s $v\cos\beta$ and $v\sin\beta$.
The physical sector contain charged scalars $H^\pm$ and three
neutral scalars $h_i$ with no definite CP parity, in the CP
conserving case these $h_i$ become scalars  $h$ and $H$ (with
$M_h<M_H$) and pseudoscalar $A$.

Generally, 2HDM permits to have relatively strong \noCP and large
FCNC. To make these effects naturally weak, the terms in Higgs
potential, giving $(\phi_1,\,\phi_2)$ mixing should be relatively
small, and the properties of the observed Higgs boson are close to
those of $h$ or $H$. In this case  masses $M_H$, $M_A$ and
$M_{H^\pm}$ are $\le 3$ TeV due to perturbativity constraint
\cite{GKO1}.

The SM--like scenario means that the squared coupling constants
(not coupling constants themselves) are close to the SM value. In
the 2HDM it can be realized in many ways even in the CP conserving
case (Table~2). The widely discussed decoupling limit corresponds
to solution $A_{h+}$ supplemented with a demand of  unnatural
strong ($\phi_1,\,\phi_2$) mixing (giving heavy $H$, $A$ and
$H^\pm$ with $M_H\approx M_A\approx M_{H^\pm}$ without
perturbativity limitation).
\begin{table}[bht]
\begin{center}
\begin{tabular}{||c|c|c|c|c|c|c||}
\noalign{\vspace{-8.5pt}} \hline\hline
&&observed&&\multicolumn{2}{|c|}{}&\\
 type &notation &Higgs
&$\chi_V$&\multicolumn{2}{|c|}{$\tan\beta$}&constraint\\
&&boson&&\multicolumn{2}{|c|}{}&\\ \hline
  &$A_{h+}$&h
&$\approx+1$&&$\lessgtr 1$&\\ \cline{2-4}\cline{6-6}
 $A_{\phi\pm}$ &$A_{H+}$&H
&$\approx+1$ &&$\lessgtr 1$&\\ \cline{2-4}\cline{6-6}
$\chi_V\approx\chi_u\approx\chi_d$&$A_{h-}$& h&$\approx
-1$&$\sqrt{\left|\frac{\epsilon_d}{\epsilon_u}\right|}$ & $\ll
1$&$\epsilon_V=-\frac{\epsilon_u\epsilon_d}{2}$\\\cline{2-4}\cline{6-6}
 &$A_{H-}$&H&$\approx -
1$&&$\gg 1$&\\ \hline \vspace{-4mm}&&&&\multicolumn{2}{|c|}{} &\\
 $B_{\phi\pm d}:$&$B_{h+d}$&h&$\approx +1$&
\multicolumn{2}{|c|}{$\sqrt{\frac{2}{\epsilon_V}}\gtrsim 10$} &
$\epsilon_u=-\frac{\epsilon_V\epsilon_d}{2}$\\\cline{2-4}
 $\chi_V\approx\chi_u\approx-\chi_d$ &$B_{H\pm d}$&H&$\approx \pm 1$&
\multicolumn{2}{|c|}{} &
\\ \hline
\vspace{-4mm}  & &&&\multicolumn{2}{|c|}{} &\\
 $B_{\phi\pm u}:$
&$B_{h\pm u}$&h&$\approx \pm 1$&
\multicolumn{2}{|c|}{$\sqrt{\frac{\epsilon_V}{2}}\lesssim 0.1$}&
$\epsilon_d=-\frac{\epsilon_V\epsilon_u}{2}$\\\cline{2-4}
 $\chi_V\approx\chi_d\approx-\chi_u$  &$B_{H+u}$&H&$\approx
+1$& \multicolumn{2}{|c|}{}&
\\\hline
 \multicolumn{7}{||c||}{\vspace{-4mm}}\\ \multicolumn{7}{||c||}{
$\chi_i=\frac{g_i}{g_i^{SM}}=\pm(1-\epsilon_i)$ with}\\
\multicolumn{7}{||c||}{ $i=V(\equiv Z,\,W)$ or $i=u(\equiv t,\,
c)$ or $i=d,\ell(\equiv b,\, \tau);\quad\epsilon_V>0$,
$\epsilon_u\epsilon_d<0$ .}\\\hline\hline
\end{tabular}
\vspace{3mm} \caption{\it Allowed realizations of SM-like scenario
in the 2HDM~(II)}
\end{center}
\label{Tab2}
\end{table}

In these SM-like cases the observed Higgs boson can be either $h$
or $H$. The pseudoscalar $A$ and other scalar, $H$ or $h$, are
almost decoupled to gauge bosons and cannot be seen at $e^+e^-$ LC
in the standard processes. If mass of any of these elusive Higgs
bosons is below 350 GeV and $\tan\beta\ll 1$, it can be seen in
$\gamma\gamma\to \gamma\gamma$ process (for $h$ -- using the low
energy part of photon spectrum) (and in $e^+e^-\to t\bar{t}H$ at
$2E>2M_t+M_H$).

In any case, one can distinguish models via measurement of
two--photon width of observed SM--like Higgs boson \cite{GKO}. For
$M_{H^\pm}= 800$ GeV the ratios of the two-photon Higgs width  and
the of cross sections for $e\gamma\to e h$ process (for latter
reaction -- at $\sqrt{s}=1.5$ TeV for the left hand polarized
photons) to their SM values are shown in Fig.~3 for the natural
set of parameters of 2HDM.
\begin{figure}[thb]
\centering \epsfig{file=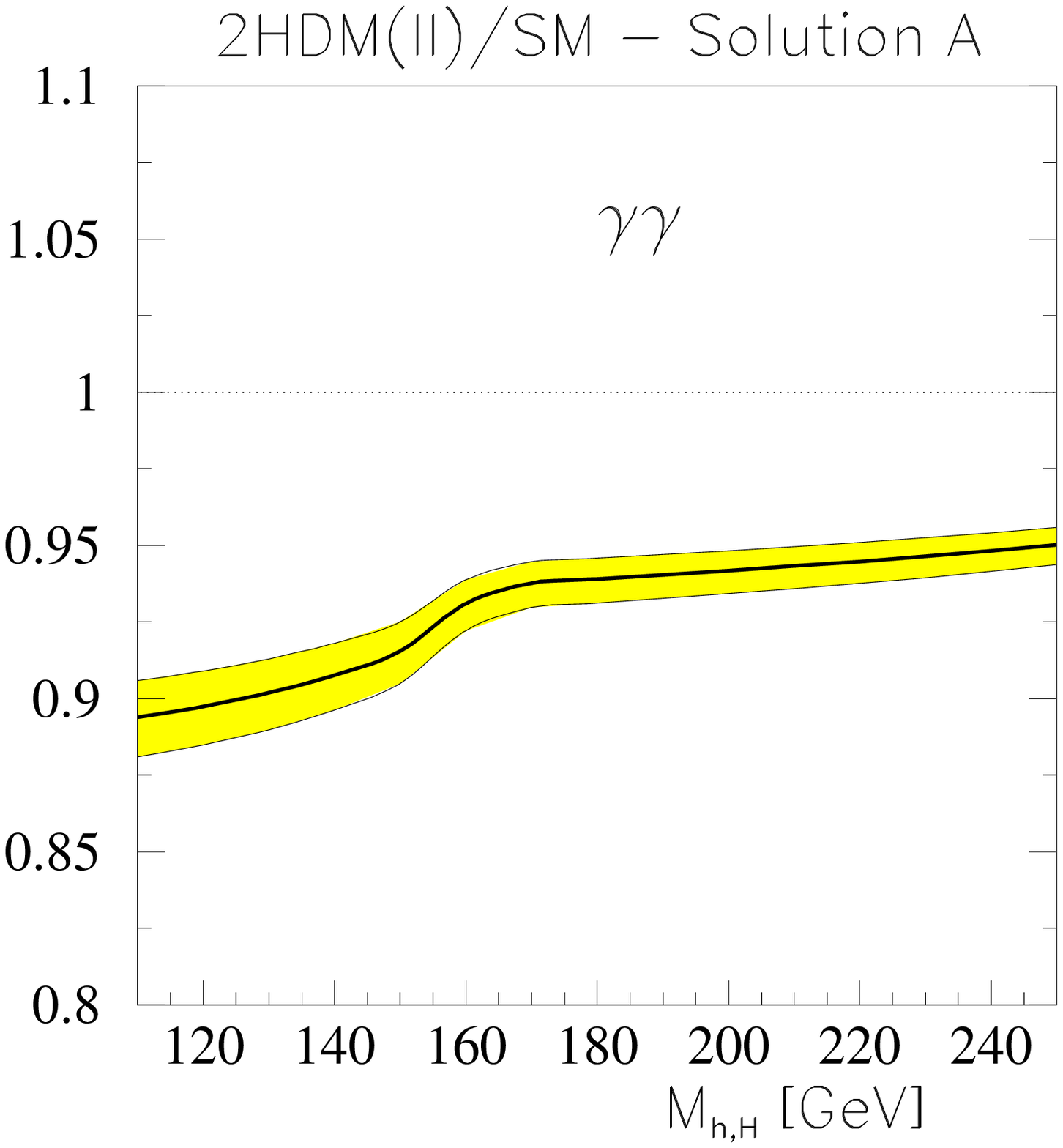,height=5cm,width=6.5cm}
 \epsfig{file=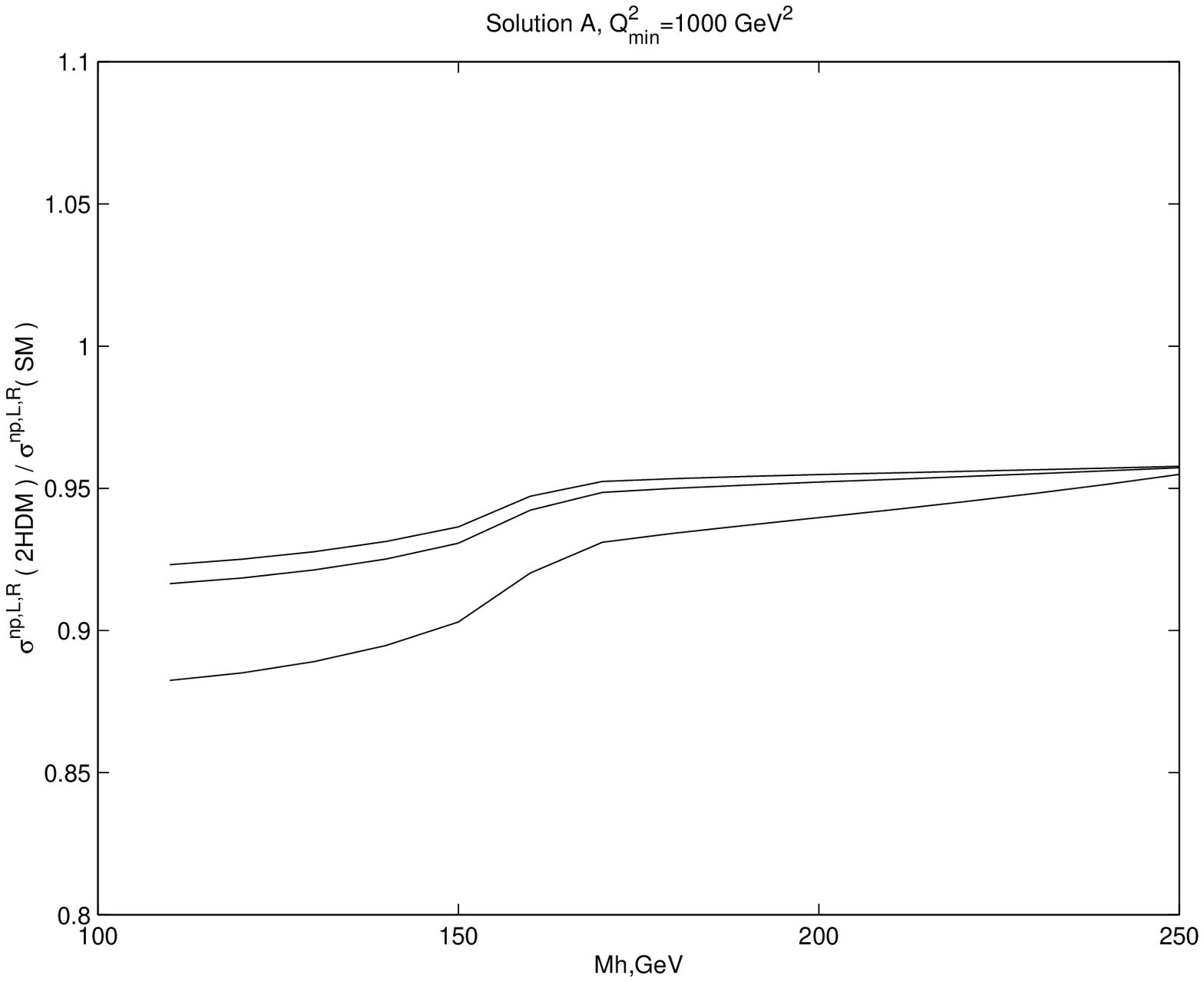,height=5cm,width=6.5cm}
\caption{\it  Solutions A and $B_{\phi\pm d}$. The ratio of
quantities in 2HDM to their SM values. The two-photon Higgs width
-- left panel; the cross section $\sigma_L(e\gamma\to eh)$ --
right panel.}
\end{figure}
The bands around central line represent possible difference of
Higgs--fermion and Higgs--$W$ couplings (squared) from their SM
values in the SM--like scenario according to anticipated
uncertainty of future measurements \cite{TESLA}.

The deviations from SM, shown in Fig.~3 for solutions $A$ and
$B_d$, are about $\sim 10$\% (one may compare with anticipated 2\%
accuracy). These deviations are given by contribution of heavy
charged Higgs bosons. For solutions $B_u$ changing of relative
sign of contributions of $t$--loop and $W$--loop increases the
observable cross section more than twice in comparison with SM.
Therefore, measurement of two-photon width at Photon Collider can
resolve these cases reliably.\\

$\blacksquare$ {\bf The possible strong interaction in the Higgs
sector}  should also be the strong interaction of longitudinal
$W$'s. It is expected to be observed in the process $\gamma\gamma
\to WW$. Based on the experience with process $\gamma\gamma
\to\pi^+\pi^-$ the following picture is expected: The strong
interaction modifies weakly the cross section near the threshold
in comparison with its SM value (but the phase of amplitude
reproduces that of strong interacting $W_LW_L$ scattering and it
can be significant). It makes strong interaction in the Higgs
sector hardly observable in the cross sections below the energies,
given by the masses of resonances, 1.5--2 TeV.

The charge asymmetry in the process $e\gamma\to eW^+W^-$ is
sensitive to this phase of amplitude even at relatively low energy
of TESLA (0.8--1 TeV) \cite{SDiego}, considerably below possible
resonance production. Indeed, the essential contribution to this
asymmetry is given by interference of two--photon production
subprocess of $WW$ pair (in C--even state) and bremsstrahlung
(one--photon) production subprocess (in C--odd state) like for the
process $e^+e^-\to e^+e^-\pi^+\pi^-$ \cite{GSS}. (The interference
with axial part of $Z$ exchange contributes additionally to this
asymmetry).

\subsection{ Anomalous interactions of gauge bosons}

At relatively low energies the New Physics cannot manifest itself
via new heavy particles. Thus, it reveals itself as certain {\em
anomalies in the interactions of known particles}. The goal of
corresponding studies is to find and discriminate these anomalies.
The correlation between coefficients of different anomalies will
be the key for understanding what is the nature of New Physics.

The practically unique process under interest in the $e^+e^-    $
mode is $e^+e^-\to WW$. With appropriate electron polarization the
neutrino exchange contribution disappears, and residual cross
section (described by photon and $Z$ boson exchange) has maximum
about 2 pb within LEP operation interval and further decreases
with energy. The cross sections of other processes with $W$
production are below 1pb at $\sqrt{s}<1$ TeV.

At Photon Collider main processes are $\gamma\gamma\to WW$,
$e\gamma\to W\nu$. Their cross sections are about 80 pb, they are
independent from energy at $\sqrt{s}>200$ GeV, what provides about
$10^7$ $W$'s per year. Due to high value of these basic cross
sections, many processes of 3-rd and 4-th order have large enough
cross sections: $e\gamma\to eWW$, $\gamma\gamma\to ZWW$,
$e\gamma\to \nu WZ$, $\gamma\gamma\to WWWW$, ..., see Fig.~4.
\begin{figure}[thb]
\centering \epsfig{file=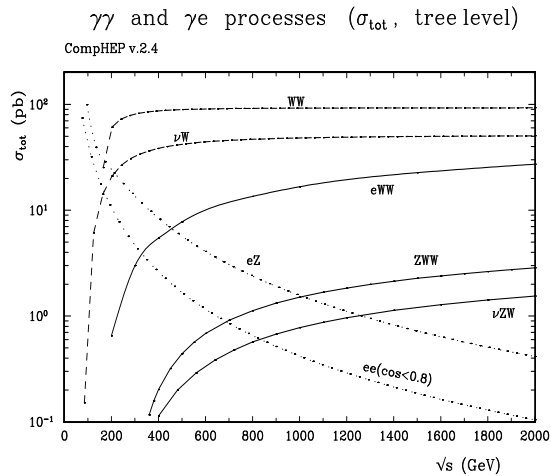,height=7cm,width=7.5cm}
\caption{\it  The 2-nd and 3-rd order processes at Photon
Collider}
\end{figure}
Large variety of these processes allows to discover and separate
well anomalies in specific processes and (or) distributions.

The cross section $e\gamma\to \nu W\propto (1-2\lambda_e)$, it is
switched on or off with variation of electron helicity
$\lambda_e$. It gives very precise test of absence of right handed
currents in the interaction of $W$ with the matter. The modern
simulations of process $e\gamma\to \nu W$  with only lepton decay
modes for $W$ decay show that the sensitivity of this reaction to
the quadruple momentum of $W$ is twice higher than it can be
reached at $e^+e^-$ LC \cite{AGP}.\\

$\blacksquare$ {\bf\boldmath The two--loop radiative corrections
to $\gamma\gamma\to WW$ and $e\gamma\to \nu W$ processes} should
be considered. They are measurable and sensitive to the (unsolved)
problems of

$\bullet$ construction of $S$--matrix in the theory with unstable
particles;

$\bullet$ gluon corrections (like Pomeron exchange) between quark
components of $W$'s.

\subsection{ QCD and Hadron Physics, t-quarks, etc. }

All the problems studied at HERA and LEP will be studied at Photon
Collider but in the wider interval of parameters and with much
higher accuracy. Among them we list those which look most
interesting now.

$\bullet$ Nature of growth of total cross sections. The widespread
concepts assume standard Regge type factorization and universal
energy behavior for different processes. With Photon Colliders, --
for the first time in particle physics at high enough energies --
one can have the set of mass shell cross sections of very high
energy processes, appropriate for the testing of factorization or
its violation. That are $\sigma_{pp}$, measured at Tevatron and
LHC, $\sigma_{\gamma p}$, measured at HERA, THERA and
$\sigma_{\gamma \gamma}$, measurable at Photon Collider. The
preliminary stage of operations with low luminosity can be used to
observe cross sections at small scattering angles.

$\bullet$ The structure function of photon is a unique QCD object
calculable completely without phenomenology impacts at high enough
photon virtuality $Q^2$ and moderate $x$ \cite{witten}. In modern
data phenomenological hadronic component of photon dominates and
accuracy of data is low. The experiments at $e\gamma$ collider
would increase the obtainable region of $Q^2$ significantly and
improve accuracy \cite{TESLAPh}.

$\bullet$ The study of charge asymmetry of produced hadrons in
$\gamma\gamma$ collisions will give quite new information about
QCD at small distances. The charge asymmetry of the produced
hadrons in the $e\gamma$ collisions with transverse momentum of
scattered electron $p_\bot\ge 30$ GeV will show in explicit form
the relation between hadron states produced by vector and axial
currents \cite{Gin}.

$\bullet$ In addition to the usually discussed problems related to
the $t$ quarks, the specific one is the study of axial anomaly in
the process $e\gamma\to et\bar{t}$ existing even in the SM. At
small transverse momenta of scattered electrons $p_\bot\to 0$ the
cross section of subprocess $\gamma Z_L\to t\bar{t}$ with
longitudinally polarized $Z$ diverges as $M_t^2/p_\bot^2$ (in
contrast to $\sigma(\gamma\gamma_L\to\bar{t}) $  tending to 0)
\cite{GIl}.

$\bullet$ The study of charge asymmetry in $e\gamma\to eb\bar{b}$
can help to discover S and D wave resonances in $b\bar{b}$.

\section{ The using of conversion region }

The $e\to\gamma$ conversion region is $e\gamma$ collider with
c.m.s. energy about 1.2 MeV and with huge luminosity about 0.1
fb$^{-1}$/sec! It will be a unique source of light Goldstone
particles (axions, majorons, etc. -- LGP, $a$) \cite{Pol}, weakly
interacting with the matter. They are expected to exist in
numerous schemes \cite{sik}.

The production processes are $ e+\gamma_0\to e+a$ and
$\gamma\gamma_0\to a$ (here $\gamma_0$ denotes laser photon).
Denoting $x=4E\omega_0/m_e^2$ and $a=m_a^2/m_e^2$, we obtain that
the LGP energy in these reactions is limited from above as
 \begin{equation}
E_a\le \frac{x+a+\sqrt{(x-a)^2-4a}}{2(x+1)}E\; \mbox{ for }
e\gamma_0\to ea\,,\;\mbox{ or }\;E_a=\frac{m_a^2}{4\omega_0}\le
\frac{x}{x+1}E\; \mbox{ for } \gamma\gamma_0\to a\,.\label{kinlim}
 \end{equation}
The angular spread of LGP is very narrow, in practice, it is given
by angular spread of incident electrons within beam ($\sim
10^{-5}$). Note that due to high density of laser photon in the
conversion region the nonlinear processes like $e+n\gamma_0\to
e+a$ $(n=2,3,..)$, etc., also become possible. The upper limit of
LGP energy in these reactions is higher than that given by
eq.~(\ref{kinlim}).

Some numerical estimates for LGP being axion or arion with mass
about 10 KeV and electron beam energy 250 GeV are presented in
Table~3 (the production rate of LGP's familon and majoron  is
negligibely low).
\begin{table}[thb]
 \centering
 \begin{tabular}{|c|c|c|c|}\hline
& $g_{aee}$ & Cross-section $\sigma (cm^2)$ & The number of
produced LGP's per year\\
 \hline  Standard axion & $2\cdot 10^{-6}$ &$2.2\cdot 10^{-35}$
 &$2\cdot 10^{10}$ \\
 \hline  "Invisible" axion &$3\cdot 10^{-8}$ &$4.9\cdot 10^{-39}$
 &$5\cdot 10^6$ \\
\hline  Arion & $2\cdot 10^{-6}$ & $2\cdot 10^{-35}$ &$2\cdot
 10^{10}$ \\\hline
  \end{tabular}
\caption{\it Estimates of typical coupling constants, production
cross sections and the number of LGP's, produced per year.}
\end{table}

To observe these LGP's, the special detector is proposed. It
should be some pin--type lead rod with radius about 2 cm and
length 300-500 m, placed in vacuum behind a shield to get rid of
the background (Fig.~5). The LGP interacts with lead nuclei like
pion but with much lower coupling constant, $a + Pb \to h$
(hadrons). The produced hadron jets with total energy $\sim
\epsilon_a$ and characteristic transverse momentum $p_\perp \sim
300$ MeV/c should be recorded in the round scinillator and
calorimeter with diameter in 1-3 m in the end of the device.
\begin{figure}[thb]
\centering \epsfig{file=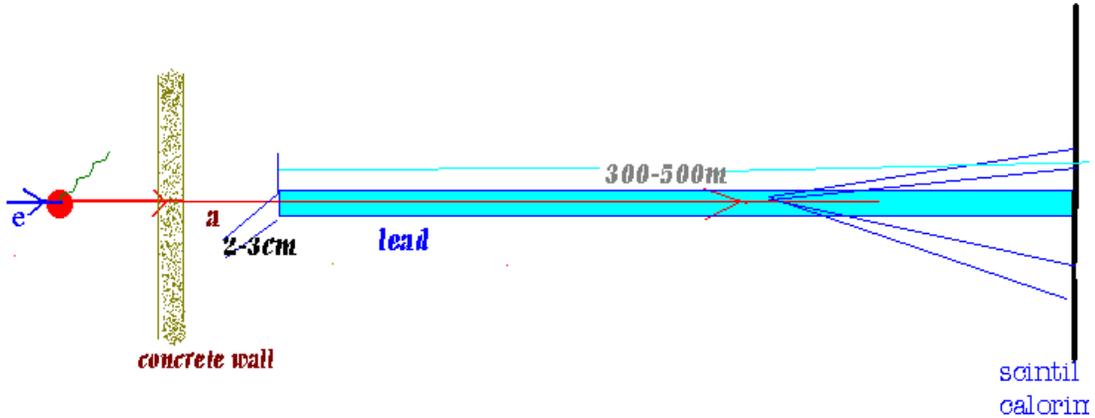,height=6cm,width=0.9\textwidth}
\caption{\it Scheme of LGP detection  }
\end{figure}

The main background is given by neutrinos produced by photons in
the shield (reactions like $\gamma+p\to \pi+...,\;\pi\to\mu\nu$).
The proposed scheme allows to reduce this background strongly.
Indeed, the energy of the main part of such neutrinos is much
lower than the upper limit (\ref{kinlim}). Besides, the produced
neutrinos are spread over the angular interval which is about 3-4
orders of magnitude wider than that for LGP's, reducing relative
flux of neutrino in the lead rod by factor $10^{-6}-10^{-8}$.

\section*{Acknowledgments}

We are thankful  D.~Anipko, V.Ilyin, I. Ivanov, M. Krawczyk,
P.~Olsen, A.~Pak, M.~Vychugin for collaboration related different
parts of paper and A.~Djouadi, V.~Serbo, M.~Spira, V.~Telnov,
P.Zerwas for useful discussions. This work was supported by RFBR
grants 99-02-17211 and 00-15-96691 and INTAS grant 00-00679.
I.F.G. is also thankful by organizers of Eriche Workshop for
invitation and support.

\end{document}